\documentclass[prl,twocolumn,showpacs,superscriptaddress,floatfix,amsmath,amssymb]{revtex4}
 
\usepackage{graphics}  
\usepackage{makeidx}
\usepackage{colordvi}
\usepackage{epsfig}

\RequirePackage{xspace}

\usepackage{relsize}
\def\babar{\mbox{\slshape B\kern-0.1em{\smaller A}\kern-0.1em
    B\kern-0.1em{\smaller A\kern-0.2em R}}}
\def\epem       {\ensuremath{e^+e^-}\xspace}
\def\qqbar {\ensuremath{q\overline q}\xspace}
\def\ccbar {\ensuremath{c\overline c}\xspace}
\def\piz   {\ensuremath{\pi^0}\xspace}
\def\pip   {\ensuremath{\pi^+}\xspace}
\def\pim   {\ensuremath{\pi^-}\xspace}

\def\Kbar  {\kern 0.2em\overline{\kern -0.2em K}{}\xspace}

\def\Km    {\ensuremath{K^-}\xspace}
\def\KS    {\ensuremath{K^0_{\scriptscriptstyle S}}\xspace}
\def\Dbar    {\kern 0.2em\overline{\kern -0.2em D}{}\xspace}

\def\Dz      {\ensuremath{D^0}\xspace}

\def\Dp      {\ensuremath{D^+}\xspace}
\def\Dm      {\ensuremath{D^-}\xspace}
\def\Dpm     {\ensuremath{D^\pm}\xspace}
\def\Dmp     {\ensuremath{D^\mp}\xspace}
\def\DpDm    {\ensuremath{\Dp {\kern -0.16em \Dm}}\xspace}
\def\Dstar   {\ensuremath{D^*}\xspace}

\def\Dstarp  {\ensuremath{D^{*+}}\xspace}
\def\Dstarm  {\ensuremath{D^{*-}}\xspace}
\def\Dstarpm {\ensuremath{D^{*\pm}}\xspace}

\def\B       {\ensuremath{B}\xspace}
\def\Bbar    {\kern 0.18em\overline{\kern -0.18em B}{}\xspace}

\def\BB      {\ensuremath{B\Bbar}\xspace}
\def\Bz      {\ensuremath{B^0}\xspace}
\def\Bzb     {\ensuremath{\Bbar^0}\xspace}
\def\BzBzb   {\ensuremath{\Bz {\kern -0.16em \Bzb}}\xspace}

\def\Y#1S{\ensuremath{\Upsilon{(#1S)}}\xspace}
\def\FourS {\Y4S}

\def\upsbb   {\ensuremath{\FourS \to \BB}\xspace}
\def\mes        {\mbox{$m_{\rm ES}$}\xspace}
\newcommand{\mev}{\ensuremath{\mathrm{\,Me\kern -0.1em V}}\xspace}
\newcommand{\gevc}{\ensuremath{{\mathrm{\,Ge\kern -0.1em V\!/}c}}\xspace}
\newcommand{\mevc}{\ensuremath{{\mathrm{\,Me\kern -0.1em V\!/}c}}\xspace}
\newcommand{\gevcc}{\ensuremath{{\mathrm{\,Ge\kern -0.1em V\!/}c^2}}\xspace}
\newcommand{\mevcc}{\ensuremath{{\mathrm{\,Me\kern -0.1em V\!/}c^2}}\xspace}
\def\to                 {\ensuremath{\rightarrow}\xspace}

\def\pep2{PEP-II}

\def\CP                {\ensuremath{C\!P}\xspace}
\def\stwob{\ensuremath{\sin\! 2 \beta   }\xspace}

\def\deltat{\ensuremath{{\rm \Delta}t}\xspace}
\def\deltamd{\ensuremath{{\rm \Delta}m_d}\xspace}
\newcommand{\progtp}    [1]  {{Prog.\ Theor.\ Phys.\ {\bf #1}}}

\begin{document}
\begin{flushleft}
\babar-PUB-07/024\\
SLAC-PUB-12506\\
\end{flushleft}

\title{\boldmath Measurement of \CP-Violating Asymmetries in $\Bz\to D^{(*)\pm}\Dmp$}

%
\author{B.~Aubert}
\author{M.~Bona}
\author{D.~Boutigny}
\author{Y.~Karyotakis}
\author{J.~P.~Lees}
\author{V.~Poireau}
\author{X.~Prudent}
\author{V.~Tisserand}
\author{A.~Zghiche}
\affiliation{Laboratoire de Physique des Particules, IN2P3/CNRS et Universit\'e de Savoie, F-74941 Annecy-Le-Vieux, France }
\author{J.~Garra~Tico}
\author{E.~Grauges}
\affiliation{Universitat de Barcelona, Facultat de Fisica, Departament ECM, E-08028 Barcelona, Spain }
% ...
\collaboration{The \babar\ Collaboration}
\noaffiliation

\begin{abstract}
We present updated measurements of \CP-violating asymmetries in the
decays $\Bz\to\Dstarpm\Dmp$ and $\Bz\to\Dp\Dm$ using $(383\pm 4)
\times 10^{6} \BB$ pairs collected by the \babar\ detector at the
PEP-II $B$ factory.  We determine the time-integrated \CP\ asymmetry
${\mathcal{A}}_{\Dstarpm\Dmp}=0.12\pm 0.06\pm 0.02$, and the
time-dependent asymmetry parameters to be $C_{\Dstarp\Dm} =0.18\pm
0.15\pm 0.04$, $S_{\Dstarp\Dm}=-0.79\pm 0.21\pm 0.06$, $C_{\Dstarm\Dp}
=0.23\pm 0.15\pm 0.04$, $S_{\Dstarm\Dp} =-0.44\pm 0.22\pm 0.06$,
$C_{\Dp\Dm} =0.11\pm 0.22\pm 0.07$, and $S_{\Dp\Dm} =-0.54\pm 0.34\pm
0.06$, where the first uncertainty is statistical and the second is
systematic.
\end{abstract}
 
  
\maketitle    

In the Standard Model (SM), \CP violation arises from a complex phase
in the Cabibbo-Kobayashi-Maskawa (CKM) quark-mixing matrix,
$V$~\cite{CKM}. Measurements of \CP asymmetries in
$\Bz\to(\ccbar)K^{(*)0}$ decays~\cite{conjugate} by the
\babar~\cite{Aubert:2002ic} and Belle~\cite{Abe:2002px} collaborations
have firmly established this effect and precisely determined the
parameter \stwob, where $\beta$ is ${\rm arg}[-V_{\rm cd}V^*_{\rm
cb}/V_{\rm td}V^*_{\rm tb}]$.  Another way to measure \stwob is to use
decays whose amplitudes are dominated by a tree-level, color-allowed
$b\to\ccbar d$ transition, such as $\Bz\to D^{(*)\pm}\Dmp$. Within the
framework of the SM, the time-dependent \CP-asymmetries of $\Bz\to
D^{(*)\pm}\Dmp$ are directly related to \stwob when corrections due to
penguin diagram contributions are neglected.  The penguin-induced
corrections have been estimated in models based on the factorization
approximation and heavy quark symmetry and are predicted to be a few
percent~\cite{Xing:1998ca,Xing:1999yx}.  However, contributions from
non-SM processes may lead to a large shift~\cite{Grossman:1996ke}.  A
significant deviation in the \stwob measurement from that of the
$\Bz\to(\ccbar)K^{(*)0}$ decays would be evidence involving new
physics beyond the SM.

Studies of the \CP violation in $b\to\ccbar d$ transitions have been
carried out by both the \babar\ and Belle collaborations. Most
recently, the Belle collaboration reported evidence of large direct
\CP violation in $\Bz\to\Dp\Dm$ where $C_{\Dp\Dm} = -0.91\pm 0.23\pm
0.06$~\cite{Fratina:2007zk}, in contradiction to the SM
expectation. However, such a large direct \CP violation has not been
observed in previous measurements with $\Bz\to D^{(*)\pm} D^{(*)\mp}$
decays, involving the same quark-level weak
decay~\cite{Aubert:2005rn,Aubert:2005hs,Aushev:2004uc,Miyake:2005qb}.

In this Letter, we present an updated measurement of \CP-violating
asymmetries in the decays $\Bz\to\Dstarp\Dm$, $\Bz\to\Dstarm\Dp$ and
$\Bz\to\Dp\Dm$.  The data used in this analysis comprise $(383\pm 4)
\times 10^{6}$ \upsbb decays collected by the \babar\ detector at the PEP-II storage
rings. The \babar\ detector is described in detail
elsewhere~\cite{Aubert:2001tu}. Monte Carlo (MC) simulation based on
GEANT4~\cite{Agostinelli:2002hh} is used to validate the analysis
procedure and to study the relevant backgrounds.

The decay rate $f_+ (f_-)$ for a neutral $B$ meson decay to a common
final state accompanied by a $\Bz(\Bzb)$ tag is given by
\begin{align}
f_\pm(\deltat) =  &{\rm e}^{ - | \deltat |/\tau_{B^0}}/4\tau_{B^0} \left\{ (1\mp\Delta w) 
\pm  (1-2w) \times \right. \nonumber \\
&\left. \left[ S\sin(\Delta m_d\deltat)
-C\cos(\Delta m_d\deltat)\right] \right\},
\label{eq:CP}
\end{align}
where $\Delta t \equiv t_{\rm rec} - t_{\rm tag}$ is the difference
between the proper decay time of the reconstructed $B$ meson ($B_{\rm
rec}$) and that of the tagging $B$ meson ($B_{\rm tag}$), $\tau_{\Bz}$
is the \Bz lifetime, and \deltamd is the difference between the heavy
and light mass eigenstates determined from the \Bz-\Bzb oscillation
frequency~\cite{Yao:2006px}.  The average mistag probability $w$
describes the effect of incorrect tags, and $\Delta w$ is the
difference between the mistag probabilities for $\Bz$ and $\Bzb$.
Since $\Dstarp\Dm$ and $\Dstarm\Dp$ are not
\CP-eigenstates, we can define a time-integrated asymmetry
$\mathcal{A}_{\Dstarpm\Dmp}$ between the rate of $\Bz\to\Dstarp\Dm$ and
$\Bz\to\Dstarm\Dp$, calculated as:
\begin{equation}
\mathcal{A}_{\Dstarpm\Dmp}=
\frac{N_{\Dstarp\Dm}-N_{\Dstarm\Dp}}
{N_{\Dstarp\Dm}+N_{\Dstarm\Dp}},
\end{equation}
where $N$ is the signal event yield.  

For $\Bz\to\Dstarpm\Dmp$, the general relations are
$S_{\Dstarpm\Dmp}=-\sqrt{1-C^2_{\Dstarpm\Dmp}}\sin(2\beta_{\rm
eff}\pm\delta)$, where $\delta$ is the strong phase difference between
$\Bz\to\Dstarp\Dm$ and $\Bz\to\Dstarm\Dp$~\cite{Aleksan:1993qk}.
Under the assumption of negligible penguin contribution, $\beta_{\rm
eff}=\beta$, $\mathcal{A}_{\Dstarpm\Dmp}= 0$ and
$C_{\Dstarp\Dm}=-C_{\Dstarm\Dp}$.  For $\Bz\to\Dp\Dm$ and in the case
of negligible penguin contribution, $C_{\Dp\Dm}$ measures direct \CP
violation and is zero, while $S_{\Dp\Dm}$ is $-\stwob$.

The selections of $\Bz\to\Dstarpm\Dmp$ and $\Bz\to\Dp\Dm$ candidates
are similar to those of our previous analysis~\cite{Aubert:2005hs}. We
reconstruct \Dstarp in its decay to $\Dz\pi^+$.  We reconstruct
candidates for \Dz and \Dp mesons in the modes $\Dz\to\Km\pip$,
$\Km\pip\piz$, $\Km\pip\pip\pim$, $\KS\pip\pim$ and
$\Dp\to\Km\pip\pip$, $\KS\pip$. We reconstruct $\Bz\to\Dp\Dm$
candidates only through the decay $\Dpm\to K^{\mp}\pi^{\pm}\pi^{\pm}$.
We require the reconstructed masses of the \Dz and \Dp candidates to
be within 20\,\mevcc of their respective nominal
masses~\cite{Yao:2006px}, except for the $\Dz\to\Km\pip\piz$
candidate, where we use a looser requirement of 40\,\mevcc.  We apply
a mass-constrained fit to the selected \Dz and \Dp candidates and
combine \Dz candidates with a \pip track, with momentum
below 450\,\mevc in the \FourS frame, to form \Dstarp candidates.

We reconstruct the \KS candidates from two oppositely charged tracks
with an invariant mass within 20\,\mevcc of the nominal \KS
mass~\cite{Yao:2006px}. The $\chi^2$ probability of the track vertex
fit must be greater than $0.1\,\%$.  We require charged kaon
candidates to be identified as such using a likelihood technique based
on the Cherenkov angle measured by the Cherenkov detector and the
ionization energy loss measured by the charged-particle tracking
systems~\cite{Aubert:2001tu}.  We form neutral pion candidates from
two photons detected in the electromagnetic
calorimeter~\cite{Aubert:2001tu}, each with energy above 30\,\mev. The
invariant mass of the pair must be within 30\,\mevcc of the nominal
\piz mass~\cite{Yao:2006px}, and we require their summed energy to be
greater than 200\,\mev. In addition, we further apply a
mass-constrained fit to the \piz candidates.

To suppress the $\epem\to\qqbar \;(q=u,d,s,\,{\rm and}\; c)$ continuum
background, we exploit the contrast between the spherical shape of
\BB\ events and the more jet-like nature of continuum events.  We
require the ratio of the second to the zeroth order Fox-Wolfram
moments~\cite{Fox:1978vu} to be less than 0.6.  We also use a Fisher
discriminant, constructed as an optimized linear combination of 11
event shape variables~\cite{Asner:1995hc}: the momentum flow in nine
concentric cones around the thrust axis of the reconstructed $\Bz$
candidate, the angle between that thrust axis and the beam axis, and
the angle between the line-of-flight of the $\Bz$ candidate and the
beam axis.  In addition, we employ a combined $D$ flight-length
significance variable, derived from the sum of flight lengths of the
two $D$ candidates~\cite{Aubert:2006ia}, to reduce background.

For each $\Bz\to D^{(*)\pm}\Dmp$ candidate, we construct a likelihood
function $\mathcal{L}_{\rm{mass}}$ from the masses and mass
uncertainties of the $D$ and \Dstar candidates~\cite{Aubert:2006ia}.
The $D$ mass resolution is modeled by a Gaussian whose variance is
determined on a candidate-by-candidate basis from its mass uncertainty
before the mass-constrained fit.  The \Dstar-$D$ mass difference
resolution is modeled by the sum of two Gaussian distributions whose
parameters are determined from simulated events.  The values of
$\mathcal{L}_{\rm{mass}}$ and $\Delta E\equiv E_B^*-E_{\rm{Beam}}$,
the difference between the \Bz candidate energy $E_B^*$ and the beam
energy $E_{\rm{Beam}}$ in the \FourS frame, are used to reduce the
combinatoric background. From the simulated events, we optimize the
maximum allowed values of $-\ln\mathcal{L}_{\rm{mass}}$ and $|\Delta
E|$ for each individual final state to obtain the highest expected
signal significance.

We extract the signal yield from the events satisfying the selection
criteria using the energy-substituted mass, $m_{\rm{ES}}\equiv
\sqrt{E^2_{\rm{Beam}}-p^{*2}_B}$, where $p^*_B$ is the $B^0$ candidate
momentum in the \FourS frame. We select the \Bz candidates that have
$m_{\rm{ES}}\ge5.23\,\gevcc$.  On average, we have $1.5$ and $1.1$ \Bz
candidates per event for $\Bz\to\Dstarpm\Dmp$ and $\Bz\to\Dp\Dm$
respectively. If more than one candidate is reconstructed in an event,
we select the candidate with the smallest value of
$-\ln\mathcal{L}_{\rm{mass}}$.  Studies using MC samples show that
this procedure results in the selection of the correct \Bz candidate
more than 95\,\% of the time.

\begin{figure*}[bth]
\epsfig{figure=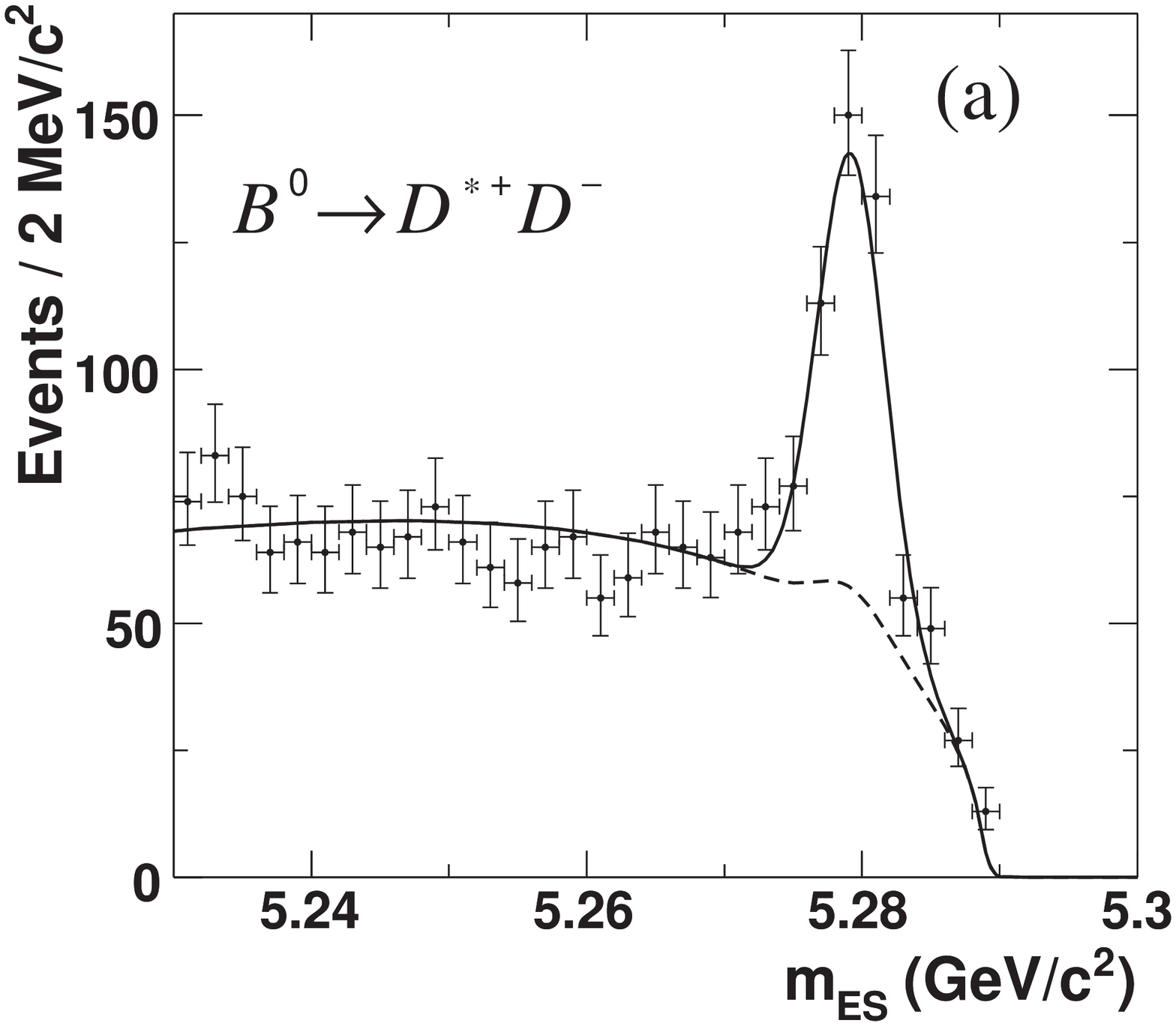,width=0.32\linewidth}
\epsfig{figure=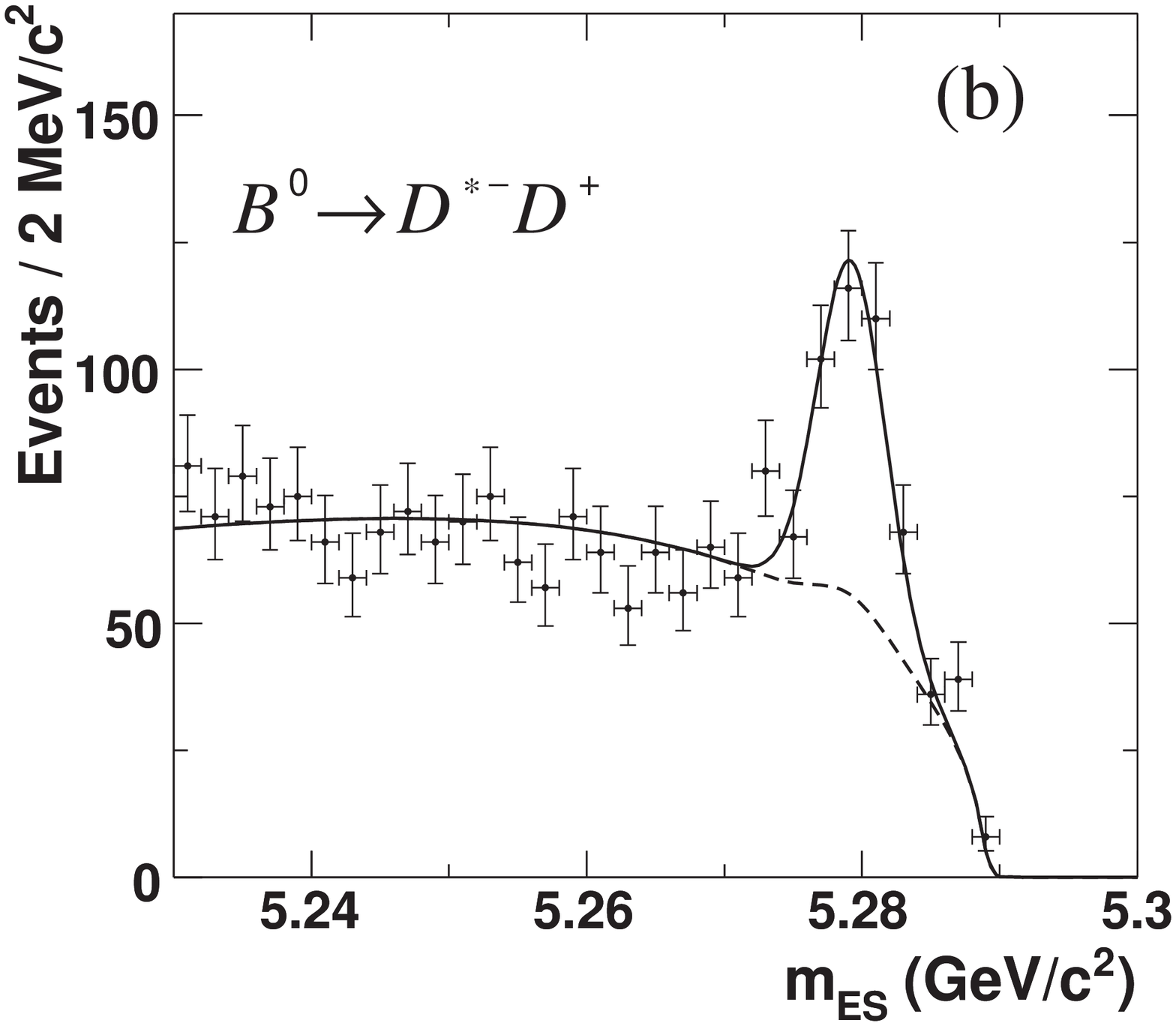,width=0.32\linewidth}
\epsfig{figure=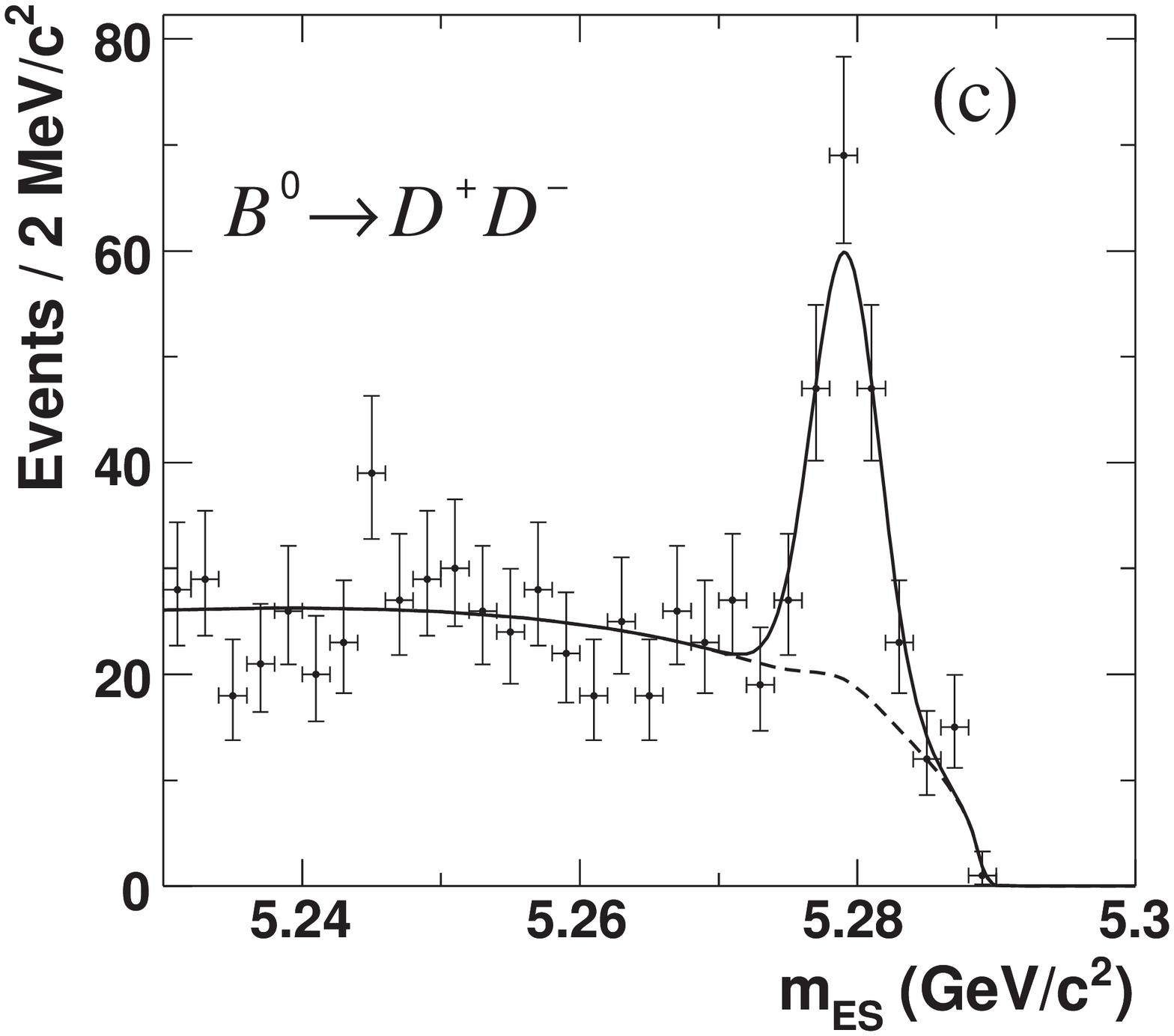,width=0.32\linewidth}
\caption{\label{fig:mesplots}Measured distribution of $\mes$ for (a)
$\Bz\to\Dstarp\Dm$, (b) $\Bz\to\Dstarm\Dp$ and (c) $\Bz\to\Dp\Dm$
candidates. The solid line is the projection of the fit result and the
dotted line represents the background components.  }
\end{figure*}

We perform an unbinned maximum likelihood fit to the $\mes$ and
\deltat\ distributions to extract the \CP\ asymmetries.  We fit the
events from $\Bz\to\Dstarp\Dm$ and $\Bz\to\Dstarm\Dp$ decays
simultaneously. The probability density function (PDF) of the \mes\
distribution consists of a Gaussian for the signal and a threshold
function~\cite{Albrecht:1990cs} for the combinatorial background.  We
expect some background events to peak in the \mes\ signal region due
to cross feed from other decay modes. We estimate the fraction of
events in the signal Gaussian due to this peaking background to be
$(8.8\pm4.4)\,\%$ for $\Bz\to\Dstarpm\Dmp$ and $(4.8\pm7.4)\,\%$ for
$\Bz\to\Dp\Dm$ using detailed MC simulations of inclusive \B\ decays.

The technique used to fit the \deltat distribution is analogous to
that used in previous \babar\ measurements described in
Ref.~\cite{Aubert:2002rg,Aubert:2004zt}.  We use information from the 
other \B meson in the event to tag the flavor of the fully reconstructed
$\Bz\to D^{(*)\pm}\Dmp$ candidate~\cite{Aubert:2002rg}.
The signal $\Delta t$ PDF in
Eq.~\ref{eq:CP} is convolved with an empirical $\Delta t$ resolution
function~\cite{Aubert:2002rg}.  The \deltat is calculated from the
measured separation $\Delta z$ between the decay vertices of $B_{\rm
rec}$ and $B_{\rm tag}$ along the collision ($z$)
axis~\cite{Aubert:2002rg}.  The $B_{\rm tag}$ decay vertex is
determined by fitting charged tracks not belonging to the $B_{\rm
rec}$ candidate to a common vertex, employing constraints from the
beam spot location and the $B_{\rm rec}$
momentum~\cite{Aubert:2002rg}.  Only events with a $\Delta t$
uncertainty less than $2.5\,\mbox{ps}$ and a measured $|\Delta t|$
less than $20\,\mbox{ps}$ are accepted for the fit to the \deltat\
distribution.  Both the signal mistag probability and the $\Delta t$
resolution function are determined from a large sample of neutral $B$
decays to flavor eigenstates, $B_{\rm flav}$.  The combinatoric 
background $\Delta t$ distributions are parameterized with an 
empirical description that includes zero and non-zero lifetime 
components~\cite{Aubert:2002rg}.
The non-zero lifetime background is allowed to have effective \CP
asymmetries, and these float in the likelihood fit. By default, we assume
that the peaking backgrounds have the same $\Delta t$ PDF as the
signal but zero \CP\ asymmetries.

The fits to the data yield $280\pm 19$ signal events for
$\Bz\to\Dstarp\Dm$, $219\pm 18$ signal events for $\Bz\to\Dstarm\Dp$,
and $131\pm 14$ signal events for $\Bz\to\Dp\Dm$, where the quoted
uncertainties are statistical only. In the region of 
$m_{\rm{ES}}> 5.27\,\gevcc$,
the signal purity is approximately 41\,\% for
$\Bz\to\Dstarp\Dm$, 34\,\% for $\Bz\to\Dstarm\Dp$,
and 46\,\% for $\Bz\to\Dp\Dm$.
The fitted \CP violating parameters are
\begin{align}
\mathcal{A}_{\Dstarpm\Dmp} &= \;\;\;0.12\pm 0.06 \pm 0.02  \nonumber \\
C_{\Dstarp\Dm} &= \;\;\;0.18\pm 0.15 \pm 0.04  \nonumber \\
S_{\Dstarp\Dm} &= -0.79\pm 0.21 \pm 0.06  \nonumber \\
C_{\Dstarm\Dp} &= \;\;\;0.23\pm 0.15 \pm 0.04  \nonumber \\
S_{\Dstarm\Dp} &= -0.44\pm 0.22 \pm 0.06  \nonumber \\
C_{\Dp\Dm} &= \;\;\;0.11\pm 0.22 \pm 0.07  \nonumber \\
S_{\Dp\Dm} &= -0.54\pm 0.34 \pm 0.06 \,,
\end{align}
where the first uncertainty is statistical and the second is
systematic.

\begin{figure*}[!htb]
\begin{center}
\epsfig{figure=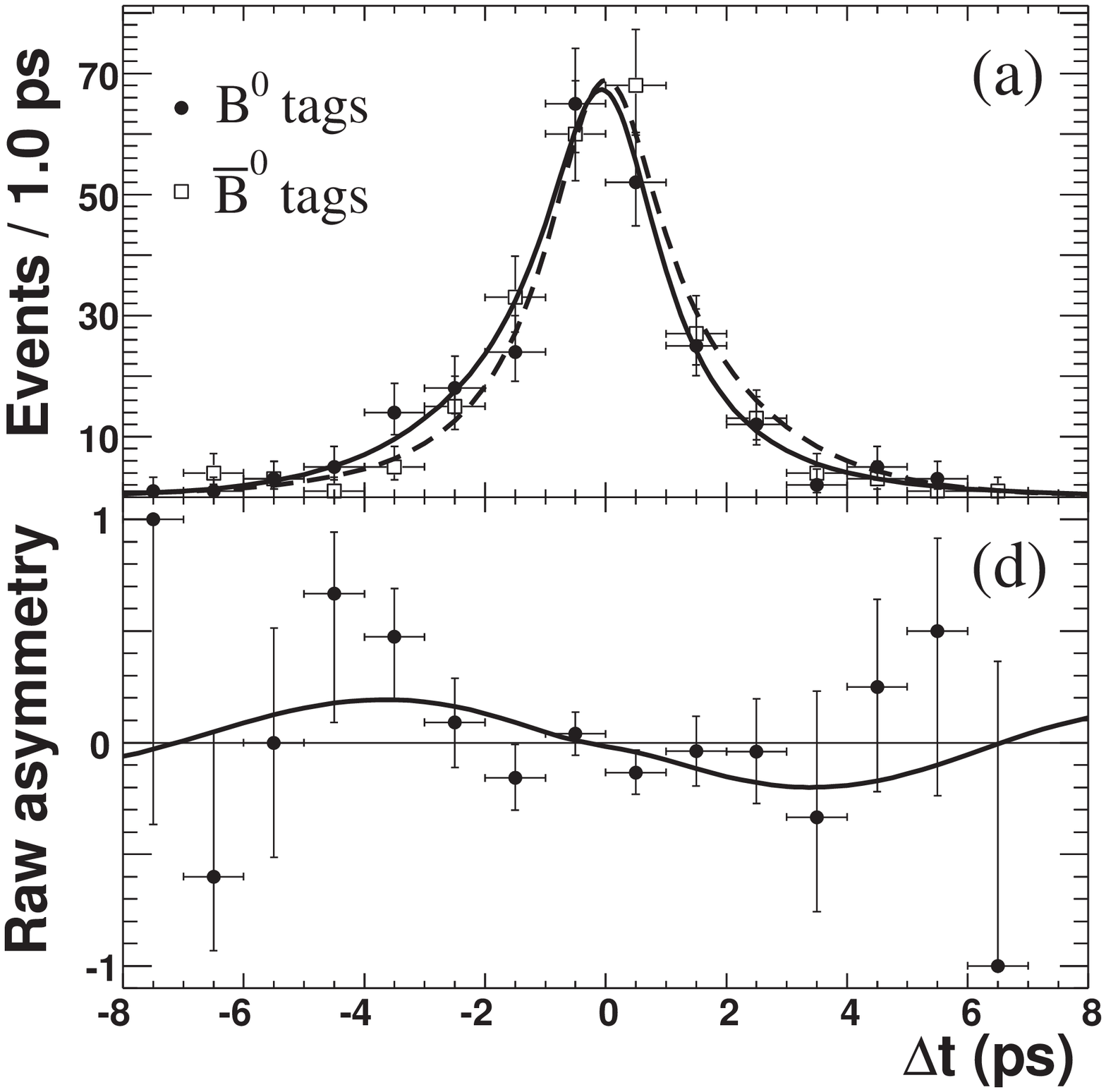,width=0.32\linewidth}
\epsfig{figure=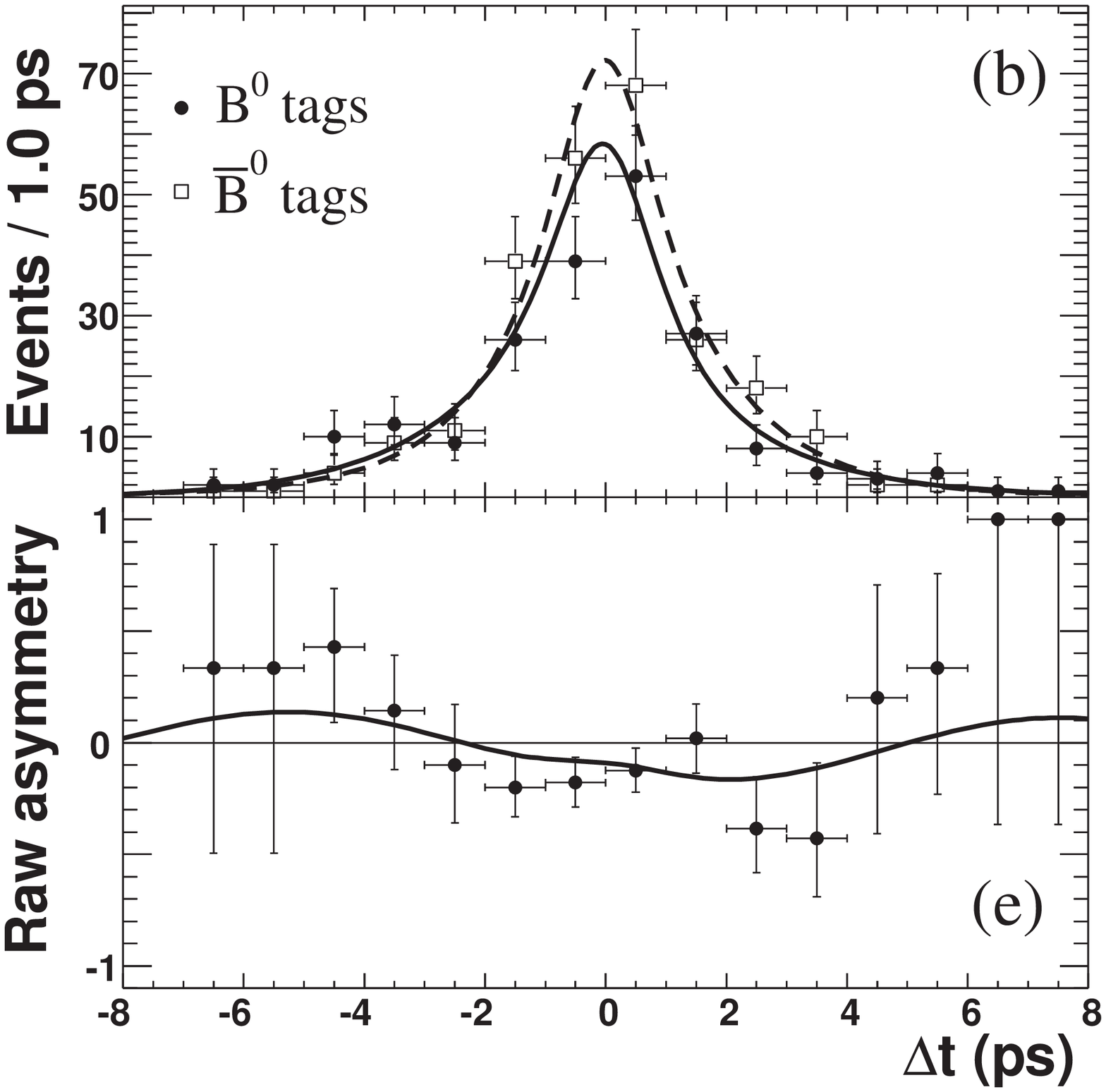,width=0.32\linewidth}
\epsfig{figure=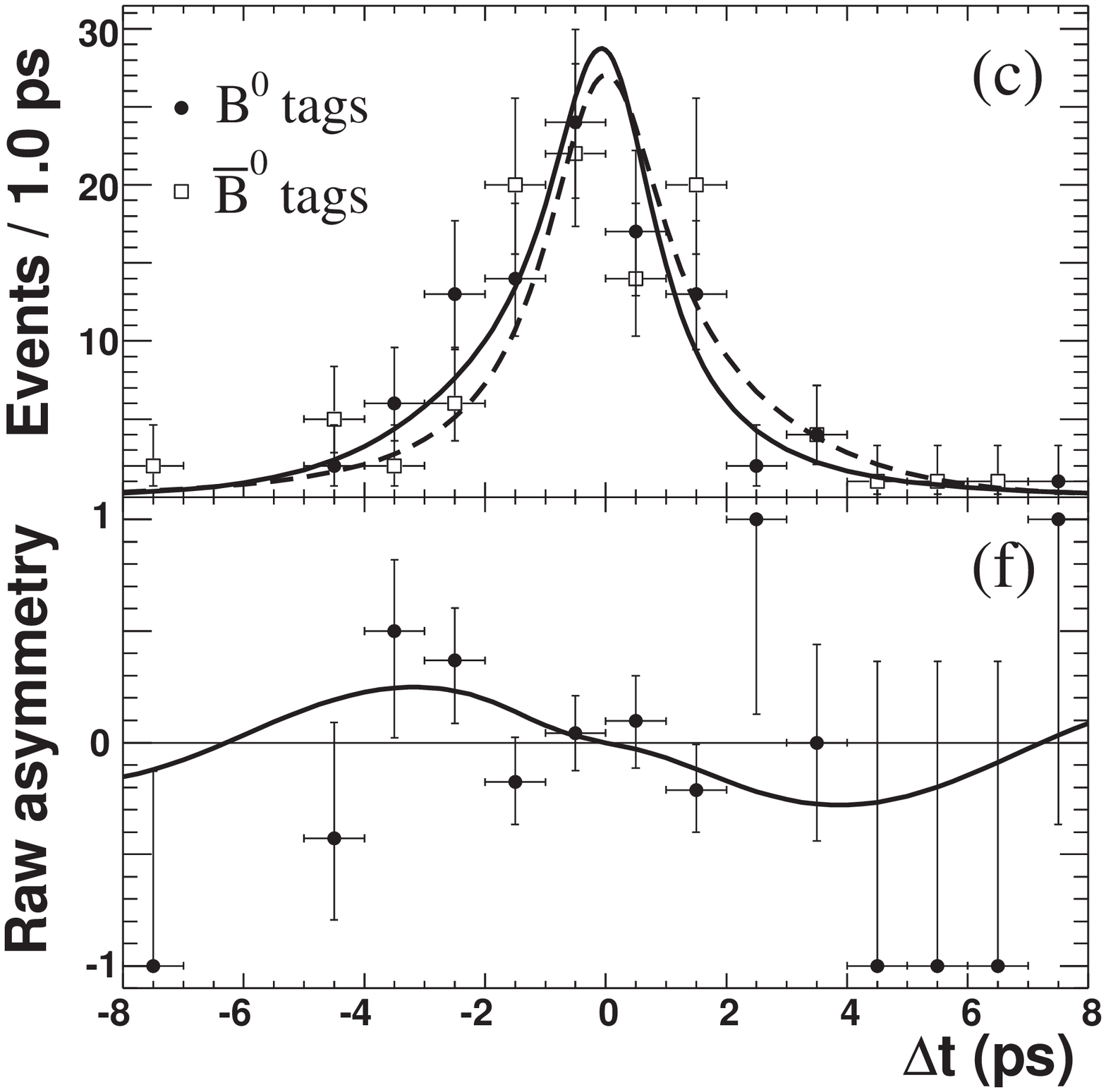,width=0.32\linewidth}
\end{center}
\caption{The distributions of \deltat and fit projections for
$\Bz\to\Dstarp\Dm$ (left), $\Bz\to\Dstarm\Dp$ (middle) and
$\Bz\to\Dp\Dm$ (right) candidates in the signal region $\mes >
5.27\,\gevcc$ with a \Bz\ or \Bzb\ tag (a)-(c).  The raw time-dependent
asymmetries $(N_{\Bz}-N_{\Bzb})/(N_{\Bz}+N_{\Bzb})$ as functions of
\deltat\ are also shown (d)-(e).  }
\label{fig:dtplots}
\end{figure*}

Projections of the fits onto \mes\ for the three different samples are
shown in Figure~\ref{fig:mesplots}.  Figure~\ref{fig:dtplots} shows
the $\Delta t$ distributions and asymmetries in yields between events
with $B^0$ and $\Bzb$ tags, overlaid with the projection of the
likelihood fit result. As a cross check, we repeat the fit by allowing
the \Bz lifetime to float. The obtained lifetime is in good
agreement with its world average~\cite{Yao:2006px}.

The systematic uncertainty of the time-integrated \CP-asymmetry
$\mathcal{A}_{\Dstarpm\Dmp}$ is dominated by the potential differences
in the reconstruction efficiencies of the positively and negatively
charged tracks ($0.014$). Other sources that contribute to the
systematic error include the estimate of the peaking background
fraction ($<0.001$), the uncertainty in the \mes\ resolution for the
$\Bz\to\Dstarpm\Dmp$ signal events ($0.005$), and a possible fit
bias ($0.004$).

The systematic uncertainties on $C$ and $S$ are evaluated separately
for each of the decay modes.  Their sources and estimates are
summarized in Table~\ref{tab:systematics}.  The systematic
uncertainties arise from the amount of possible background that tends
to peak under the signal and its \CP asymmetry, the assumed
parameterization of the $\Delta t$ resolution function, the possible
differences between the $B_{\rm flav}$ and signal mistag fractions,
the knowledge of the event-by-event beam-spot position, the
uncertainties from the finite MC sample used, the possible
interference between the suppressed $\bar{b}\to\bar{u}c\bar{d}$ and
the favored $b\to c\bar{u}d$ amplitudes in some tag-side
decays~\cite{Long:2003wq}, and the uncertainty in the \mes\ resolution 
for the signal events.  All of the systematic
uncertainties are found to be much smaller than the statistical
uncertainties.

\begin{table*}[!htb]
\begin{ruledtabular}
\begin{tabular}{lcccccc}
Source  &  $C_{\Dstarp\Dm}$ & $S_{\Dstarp\Dm}$ & $C_{\Dstarm\Dp}$ & 
$S_{\Dstarm\Dp}$ & $C_{\Dp\Dm}$ & $S_{\Dp\Dm}$ 
\\ \hline
Peaking backgrounds                  & 0.026 & 0.041 & 0.027 & 0.031 & 0.044 & 0.042  \\
\deltat\ resolution parameterization & 0.011 & 0.021 & 0.013 & 0.012 & 0.015 & 0.020 \\
Mistag fraction differences          & 0.014 & 0.011 & 0.016 & 0.012 & 0.023 & 0.013  \\
Beam-spot position                   & 0.004 & 0.006 & 0.007 & 0.036 & 0.005 & 0.002 \\
$\deltamd$, $\tau_B$                 & 0.002 & 0.003 & 0.003 & 0.004 & 0.001 & 0.004  \\
MC statistics                        & 0.011 & 0.015 & 0.011 & 0.015 & 0.036 & 0.023  \\
Tag-side interference and others     & 0.016 & 0.025 & 0.017 & 0.020 & 0.020 & 0.013 
\\ \hline
Total                                & 0.037 & 0.056 & 0.040 & 0.056 & 0.066 & 0.055 \\
\end{tabular}
\end{ruledtabular}
\caption{ Sources of systematic error on time-dependent \CP\ asymmetry
parameters for the decays $\Bz\to\Dstarpm\Dmp$ and $\Bz\to\Dp\Dm$.}
\label{tab:systematics}
\end{table*}

Since $\Dstarp\Dm$ and $\Dstarm\Dp$ are not \CP-eigenstates, it is
also illustrative to express the measured \CP-violating parameters $C$
and $S$ in a slightly different parametrization~\cite{Aubert:2003wr}:
$C_{\Dstar D}=(C_{\Dstarp\Dm}+C_{\Dstarm\Dp})/2$, $\Delta C_{\Dstar
D}=(C_{\Dstarp\Dm}-C_{\Dstarm\Dp})/2$, $S_{\Dstar
D}=(S_{\Dstarp\Dm}+S_{\Dstarm\Dp})/2$ and $\Delta S_{\Dstar
D}=(S_{\Dstarp\Dm}-S_{\Dstarm\Dp})/2$.  The quantities $C_{\Dstar D}$
and $S_{\Dstar D}$ parametrize flavor-dependent direct \CP violation,
and mixing-induced \CP violation related to the angle $\beta$,
respectively.  The parameters $\Delta C_{\Dstar D}$ and $\Delta
S_{\Dstar D}$ are insensitive to \CP violation. $\Delta C_{\Dstar D}$
describes the asymmetry between the rates
$\Gamma(\Bz\to\Dstarp\Dm)+\Gamma(\Bzb\to\Dstarm\Dp)$ and
$\Gamma(\Bz\to\Dstarm\Dp)+\Gamma(\Bzb\to\Dstarp\Dm)$, while $\Delta
S_{\Dstar D}$ is related to the strong phase difference, $\delta$. We
find
\begin{align}
C_{\Dstar D} &= \;\;\;0.21\pm 0.11 \pm 0.03  \nonumber \\
S_{\Dstar D} &= -0.62\pm 0.15 \pm 0.04  \nonumber \\
\Delta C_{\Dstar D} &= -0.02\pm 0.11 \pm 0.03  \nonumber \\
\Delta S_{\Dstar D} &= -0.17\pm 0.15 \pm 0.04  \,,
\end{align}
where the first uncertainty is statistical and the second is
systematic.

In summary, this letter reports updated measurements of the \CP
violating asymmetries for the decays $\Bz\to\Dstarpm\Dmp$ and
$\Bz\to\Dp\Dm$.  These measurements supersede the previous \babar\
results~\cite{Aubert:2005hs}, with a more than $50\,\%$ reduction in
the statistical uncertainties.  The time-dependent asymmetries are
consistent with the SM predictions within their statistical
uncertainties. We do not see evidence of large direct \CP\ violation in
the decay $\Bz\to\Dp\Dm$ as reported by the Belle
Collaboration~\cite{Fratina:2007zk}.

We are grateful for the excellent luminosity and machine conditions
provided by our \pep2\ colleagues,
and for the substantial dedicated effort from
the computing organizations that support \babar.
The collaborating institutions wish to thank
SLAC for its support and kind hospitality.
This work is supported by
DOE
and NSF (USA),
NSERC (Canada),
CEA and
CNRS-IN2P3
(France),
BMBF and DFG
(Germany),
INFN (Italy),
FOM (The Netherlands),
NFR (Norway),
MIST (Russia),
MEC (Spain), and
STFC (United Kingdom).
Individuals have received support from the
Marie Curie EIF (European Union) and
the A.~P.~Sloan Foundation.


\begin{thebibliography}{99}

\bibitem{CKM}
\hyphenation{Ko-ba-ya-shi}
N.~Cabibbo, \prl {\bf 10}, 531 (1963); M.~Kobayashi and T.~Maskawa, \progtp {\bf 49}, 652 (1973).

\bibitem{conjugate}
We imply charge conjugate modes throughout the paper.

\bibitem{Aubert:2002ic}
\babar\ Collaboration, B.~Aubert {\it et al.}, 
Phys.\ Rev.\ Lett.\  {\bf 89}, 201802 (2002).

\bibitem{Abe:2002px}
Belle Collaboration, K.~Abe {\it et al.}, 
Phys.\ Rev.\ D {\bf 66}, 071102 (2002).

\bibitem{Xing:1998ca}
Z.~Z.~Xing,
Phys.\ Lett.\  B {\bf 443}, 365 (1998).

\bibitem{Xing:1999yx}
Z.~Z.~Xing,
Phys.\ Rev.\ D {\bf 61}, 014010 (2000).

\bibitem{Grossman:1996ke}
Y.~Grossman and M.~P.~Worah,
Phys.\ Lett.\  B {\bf 395}, 241 (1997).

\bibitem{Fratina:2007zk}
Belle Collaboration, S.~Fratina {\it et al.},
Phys.\ Rev.\ Lett.\  {\bf 98}, 221802 (2007).

\bibitem{Aubert:2005rn}
\babar\ Collaboration, B.~Aubert {\it et al.}, 
Phys.\ Rev.\ Lett.\  {\bf 95}, 151804 (2005).

\bibitem{Aubert:2005hs}
\babar\ Collaboration, B.~Aubert {\it et al.}, 
Phys.\ Rev.\ Lett.\  {\bf 95}, 131802 (2005).

\bibitem{Aushev:2004uc}
Belle Collaboration, T.~Aushev {\it et al.}, 
Phys.\ Rev.\ Lett.\  {\bf 93}, 201802 (2004).

\bibitem{Miyake:2005qb}
Belle Collaboration, H.~Miyake {\it et al.}, 
Phys.\ Lett.\  B {\bf 618}, 34 (2005).

\bibitem{Aubert:2001tu}
\babar\ Collaboration, B.\ Aubert {\em et al.}, 
Nucl.\ Instr.\ Meth.\ Phys. Res., Sect. A {\bf 479}, 1 (2002).

\bibitem{Agostinelli:2002hh}
GEANT4 Collaboration, S.~Agostinelli {\it et al.},
Nucl.\ Instr.\ Meth.\ Phys. Res., Sect. A {\bf 506}, 250 (2003).

\bibitem{Yao:2006px}
W.~M.~Yao {\it et al.}  (Particle Data Group),
J.\ Phys.\ G {\bf 33} (2006) 1.

\bibitem{Aleksan:1993qk}
R.~Aleksan, A.~Le Yaouanc, L.~Oliver, O.~Pene and J.~C.~Raynal,
Phys.\ Lett.\  B {\bf 317}, 173 (1993).

\bibitem{Fox:1978vu}
G.~C.~Fox and S.~Wolfram,
Phys.\ Rev.\ Lett.\  {\bf 41}, 1581 (1978).

\bibitem{Asner:1995hc}
CLEO Collaboration, D.~M.~Asner {\it et al.},
Phys.\ Rev.\  D {\bf 53}, 1039 (1996).

\bibitem{Aubert:2006ia}
\babar\ Collaboration, B.~Aubert {\it et al.},
Phys.\ Rev.\  D {\bf 73}, 112004 (2006).

\bibitem{Albrecht:1990cs}
ARGUS Collaboration, H.~Albrecht {\it et al.},
Z.\ Phys.\ C {\bf 48}, 543 (1990).

\bibitem{Aubert:2002rg}
\babar\ Collaboration, B.\ Aubert {\em et al.},
Phys.\ Rev.\ D {\bf 66}, 032003 (2002).

\bibitem{Aubert:2004zt}
\babar\ Collaboration, B.~Aubert {\it et al.},
Phys.\ Rev.\ Lett.\  {\bf 94}, 161803 (2005).

\bibitem{Long:2003wq}
O.~Long, M.~Baak, R.~N.~Cahn, and D.~Kirkby,
Phys.\ Rev.\ D {\bf 68}, 034010 (2003).

\bibitem{Aubert:2003wr}
\babar\ Collaboration, B.~Aubert {\it et al.},
Phys.\ Rev.\ Lett.\  {\bf 91}, 201802 (2003).

\end{thebibliography}
\end{document}